\documentclass[12pt,aps,prb,preprint]{revtex4}   

\usepackage{amsmath}    
\usepackage{graphicx}   

\begin{document}

\begin{flushright}
DO-TH 10/12
\end{flushright}

\title{The optimal angle of Release in Shot Put}
\author{Alexander Lenz}
\affiliation{Institut f{\"u}r Physik, Technische Universit{\"a}t Dortmund, D-44221 Dortmund, Germany}

\affiliation{Institut f{\"u}r Theoretische Physik, Universit{\"a}t Regensburg, D-93040 Regensburg, Germany}
\author{Florian Rappl}
 \affiliation{Institut f{\"u}r Theoretische Physik, Universit{\"a}t Regensburg, D-93040 Regensburg, Germany}
 \email{florian.rappl@rsinet.de, alexander.lenz@physik.uni-regensburg.de}   
\date{\today}

\begin{abstract}
We determine the optimal angle of release in shot put.
The simplest model - mostly used in textbooks - gives a value of $45^\circ$,
while measurements of top athletes cluster around $37 - 38^\circ$.
Including simply the height of the athlete the theory prediction goes down
to about $42^\circ$ for typical parameters of top athletes. Taking further the correlations of the initial velocity
of the shot, the angle of release and the height of release into account we 
predict values around $37 - 38^\circ$, which coincide perfectly with 
the measurements.
\end{abstract}

\maketitle

\section{Introduction}
We investigate different effects contributing to the determination of the optimal angle of 
release at shot put. Standard text-book wisdom tells us that the optimal angle is $45^\circ$, 
while measurements of world-class athletes 
\cite{Kuhlow75,Dessureault78,McCoy84,Susanka88,Bartonietz95,Tsirakos95,Luhtanen97} typically give values 
of below $40^\circ$. 
In Table \ref{Tab1} we show  the data from the olympic games in 1972 given by Kuhlow (1975) \cite{Kuhlow75}
with an average angle of release of about $38^\circ$. The measurements of 
Dessureault                  (1978) \cite{Dessureault78},
McCoy et al.                 (1984) \cite{McCoy84},
Susanaka and Stepanek        (1988) \cite{Susanka88},
Bartonietz and Borgstr{\"o}m (1995) \cite{Bartonietz95},
Tsirakos et al.              (1995) \cite{Tsirakos95}
and
Luhtanen et al.              (1997) \cite{Luhtanen97}
give an average angle of release of about $37^\circ$.
\\
This obvious deviation triggered already considerable interest in the literature
\cite{Tricker67,Garfoot68,Zatsiorski69,Tutevich69,Hay73,Lichtenberg78,Zatsiorsky81,McWatt82,Townend84,Dyson86,Hubbard88,deMestre90,Gregor90,deMestre98,Maheras98,Yeadon98,Linthorne01,Hubbard01,Bace02,Mizera02,Cross04,Oswald04,deLuca05,Linthorne06}.
Most of these investigations obtained values below $45^\circ$ but still considerably above the measured
values. E.g. in the classical work of Lichtenberg and Wills (1976) \cite{Lichtenberg78} optimal
release angles of about $42^\circ$ were found by including the effect of the height of an athlete.
\\
We start by redoing the analysis of Lichtenberg and Wills (1976) \cite{Lichtenberg78}. 
Next we investigate the effect of air resistance. Here we find as expected \cite{Tutevich69,Lichtenberg78} 
that in the case of shot put air resistance gives a negligible contribution\footnote{Oswald and Schneebeli (2004)\cite{Oswald04}
find relatively large effects due to air resistance. We could trace back their result to an error
in the formula for the area of the shot. They used $A = d^2 \pi$ instead of $A = r^2 \pi$ ($d$ diameter, $r$ radius)
and have therefore a force that is four times as large as the correct one.}.
If the initial velocity $\vec{v}_0$, the release height $h$ and the release angle $\theta$ are known, 
the results obtained up to that point are exact. We provide a computer program to determine graphically
the trajectory of the shot for a given set of $\vec{v}_0$, $h$ and $\theta$
including air resistance and wind. 
\\
Coming back to the question of the optimal angle of release we give up the assumption of 
Lichtenberg and Wills (1976) \cite{Lichtenberg78}, that 
the initial velocity, the release height and the release angle are uncorrelated.
This was suggested earlier in the literature 
\cite{Tricker67,Hay73,Dyson86,Hubbard88,deMestre90,deMestre98,Maheras98,Linthorne01,deLuca05}.
We include three correlations:
\begin{itemize}
\item The angle dependence of the release height; this was discussed in detail by de Luca (2005) \cite{deLuca05}.
\item The angle dependence of the force of the athlete; this was suggested for javeline throw by Red and Zogaib (1977)
      \cite{Red77}. In particular a inverse proportionality between the initial velocity and the angle of release was found.
      This effect was discussed for the case of shot put in
      McWatt    (1982)\cite{McWatt82},
      McCoy     (1984)\cite{McCoy84},
      Gregor    (1990)\cite{Gregor90}
      and
      Linthorne (2001)\cite{Linthorne01}.
\item The angle dependence of the initial velocity due to the effect of gravity during the period of release;
      this was discussed e.g. in Tricker and Tricker (1967) \cite{Tricker67}, 
      Zatsiorski and Matveev (1969) \cite{Zatsiorski69}, Hay (1973) \cite{Hay73} and 
      Linthorne (2001)\cite{Linthorne01}.
\end{itemize}
To include these three correlations we still need information about the angle dependence of the force of the
athlete. In principle this has to be obtained by measurements with each invididual athlete.
To show the validity of our approach we use a simple model for the angle dependence of the force
and obtain realistic values for the optimal angle of release.
\\
Our strategy is in parts similar to the nice and extensive work of Linthorne (2001) \cite{Linthorne01}.
While Linthorne's approach is based on experimental data on $v(\theta)$ and $h(\theta)$ our approach is more theoretical.
We present some toy models that predict the relation $v \propto - \theta$ found by  Red and Zogaib (1977)
\cite{Red77}.
\\
We do not discuss possible deviations between the flight distance of the shot and the official 
distance. Here were refer the interested reader to the work of Linthorne (2001) \cite{Linthorne01}.
\section{Elementary Biomechanics of Shot Put}
\subsection{The simplest approach}
\begin{figure}[h]
\begin{center}
\scalebox{1.40}{\includegraphics{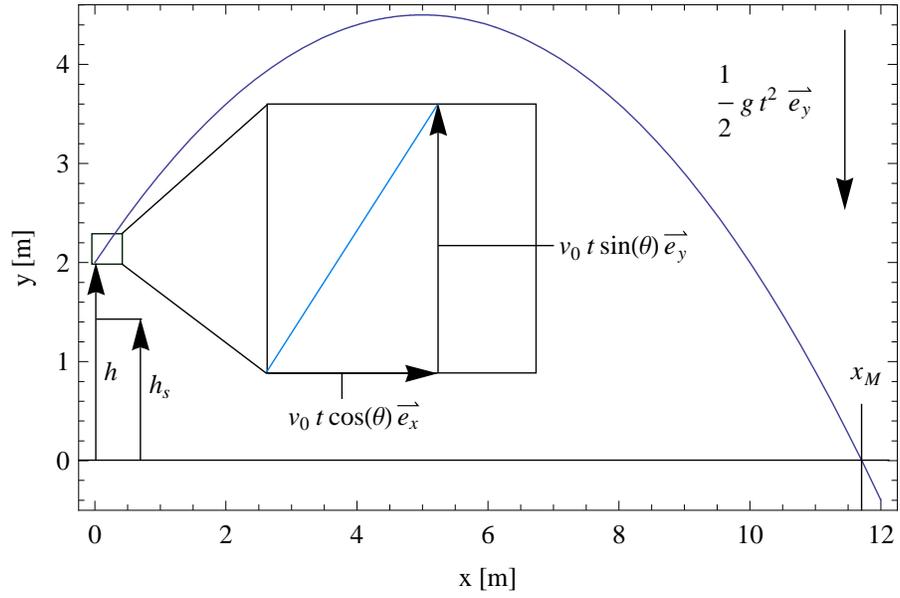}}
\caption{\label{Fig1}The setup for our calculations with the angle $\theta$ of the velocity $v_0$ split into $x$ and $y$ direction and the height of the throw $h$ with the shoulder height $h_s$ of the athlete}
\end{center}
\end{figure}

Let us start with the simplest model for shot put. The shot is
released from a horizontal plane with an initial velocity $\vec{v}_0$ 
under the angle $\theta$ relative to the plane.
We denote the horizontal distance with $x$ and the vertical
distance with $y$. The maximal height of the shot is denoted by $y_M$; 
the shot lands again on the plane after travelling the horizontal distance $x_M$, see 
Fig.\ref{Fig1}.

\noindent
Solving the equations of motions $ \vec{F} = m \dot{\vec{x}}$ 
with the initial condition
\begin{equation}
\dot{\vec{x}} (0) = \vec{v}_0 = v_0 \left( \begin{array}{l} \cos \theta \\ \sin \theta \end{array} 
\right) \, ,
\end{equation}
one obtains
\begin{eqnarray}
x(t) & = &  v_0 \cos \theta t \, ,
\\
y(t) & = &  v_0 \sin \theta t - \frac{1}{2} g t^2 \, ,
\\
\Rightarrow
y(x) & = & x \tan \theta - \frac{1}{2} \frac{g x^2}{v_0^2 \cos^2 \theta} \, .
\label{EOM1}
\end{eqnarray}
The maximal horizontal distance is obtained by setting $y(x)$ equal to zero
\begin{equation}
x_M = \frac{v_0^2}{g} \sin 2 \theta \, .
\label{xM0}
\end{equation}
From this result we can read off that the optimal angle is $\theta^{Opt.} = 45^\circ$ - 
this is the result that is obtained in many undergraduate textbooks. It is however considerably
above the measured values of top athletes. Moreover, Eq.(\ref{xM0}) shows that the maximal 
range at shot put depends quadratically on the initial velocity of the shot.
\subsection{The effect of the height of the athlete}

\begin{figure}[h]
\begin{center}
\scalebox{1.40}{\includegraphics{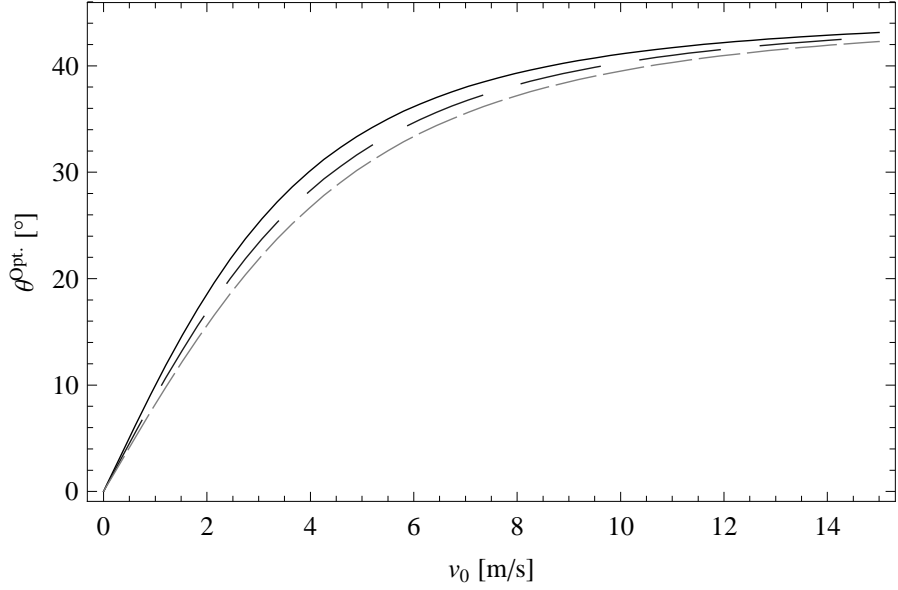}}
\caption{\label{Fig2}The optimal angle $\theta^{Opt.}$ for (from top to bottom) $h=1.6,2.0,2.4$ m in dependence of the start velocity $v_0$ taking the height of the athlete into account}
\end{center}
\end{figure}

Next we take the height of the athlete into account, this was described first in 
Lichtenberg and Wills (1976) \cite{Lichtenberg78}. 
Eq. (\ref{EOM1}) still holds for that case. We denote the height at which the shot is released
with $h$. The maximal horizontal distance is now obtained 
by setting $y(x)$ equal to $-h$.
\begin{equation}
x_M = \frac{v_0^2 \cos \theta }{g} 
\left( \sin  \theta + \sqrt{\sin^2 \theta + \frac{2gh}{v_0^2}}
\right) \, .
\label{xM}
\end{equation} 
This equation holds exactly if the parameters $v_0$, $h$ and $\theta$ are known and if the air
resistance is neglected.
\\
Assuming that the parameters $v_0$, $h$ and $\theta$ are independent of each other
we can determine the optimal angle of release by setting the derivative of $x_M$ with respect to
$\theta$ to zero.
\begin{equation}
\sin \theta^{Opt.} = \frac{1}{\sqrt{2 \left( 1 +\frac{hg}{v_0^2} \right)}} \, .
\end{equation}
The optimal angle is now always smaller than $45^\circ$. With increasing $h$ the optimal
angle is getting smaller, therefore taller athletes have to release the shot more flat. 
The dependence on the initial velocity $v_0$ is more complicated. Larger values of $v_0$ favor larger values of $\theta$.
We show the optimal angle for three fixed values of $h=1.6$ m, $2$ m and $2.4$ m in dependence of $v_0$ in Fig.\ref{Fig2}.

\noindent
With the average values from Table \ref{Tab1} for $h= 2.15$ m and 
$v_0 \approx 13.7$ m/s we obtain an optimal angle of $\theta^{Opt.} \approx 42^\circ$,
while the average of the measured angles from Table \ref{Tab1} is about
$38^\circ$.
We conclude therefore that the effect of including the height of the athlete goes 
in the right direction, but 
the final values for the optimal angle are still larger than the measured ones. 
In our example the initial discrepancy between theory and measurement of $45^\circ - 38^\circ = 7^\circ$
is reduced to $42^\circ - 38^\circ = 4^\circ$.
These findings coincide with the ones from
Lichtenberg and Wills (1976) \cite{Lichtenberg78}.
\\
For $h \approx 2$  m and $v_0 \geq 12$ m/s ($hg/v_0^2 \leq 0.136$) we can also expand the expression 
for the optimal angle
\begin{equation}
\sin \theta^{Opt.} 
                   \approx \frac{1}{\sqrt{2}} \left[ 1 - \frac{1}{2} \frac{g h }{v_0^2} 
                  + {\cal O} \left( \frac{hg}{v_0^2} \right)^2 \right] 
                  =  \frac{1}{\sqrt{2}} \left[ 1 - \frac{1}{4} \frac{E_{pot}}{E_{kin}} 
                  + {\cal O} \left( \frac{hg}{v_0^2} \right)^2 \right] \, ,
\end{equation}
with the kinetic energy $E_{kin} = \frac12   m v_0^2$ and the potential energy $E_{pot} = mgh$. $m$ denotes
the mass of the shot.
\\
By eliminating different variables from the problem, Lichtenberg and Wills (1976) \cite{Lichtenberg78} 
derived several expressions for the maximum range at shot put:
\begin{eqnarray}
x_M & = & h \tan 2 \theta^{Opt.}                         \hspace{1.4cm} 
(v_0 \, \, \, \mbox{eliminated})\, ,
\\
\label{xMv0el}
    & = & \frac{v_0^2}{g} \sqrt{1 + \frac{2gh}{v_0^2}}   \hspace{1cm} 
(\theta^{Opt.} \, \, \, \mbox{eliminated}) \, ,
\label{xMthel}
\\  
    & = & \frac{v_0^2}{g} \cot \theta^{Opt.}             \hspace{1.4cm} 
(h \, \, \, \mbox{eliminated}) \, .
\label{xMhel}
\end{eqnarray}
Expanding the expression in Eq.(\ref{xMthel}) in $hg/v_0^2$ one gets
\begin{eqnarray}
x_M & = & \frac{v_0^2}{g} + h  + {\cal O} \left( \frac{hg}{v_0^2} \right)^2 
\\
    & = & \frac{2 E_{kin} + E_{pot}}{mg} + {\cal O} \left( \frac{hg}{v_0^2} \right)^2 \, .
\end{eqnarray}
Here we can make several interesting conclusions
\begin{itemize}
     \item To zeroth order in  $hg/v_0^2$ the maximal horizontal distance is simply given by $v_0^2/g$.
           This can also be read off from Eq. (\ref{xM0}) with $\theta = \theta^{Opt.} = 45^\circ$.
     \item To first order in  $hg/v_0^2$ the maximal horizontal distance  is $v_0^2/g + h$. Releasing
           the shot from 10 cm more height results in a 10 cm longer horizontal distance. 
     \item The kinetic energy is two times more important than the potential energy. If an athlete has the additional
           energy $\delta E$ at his disposal, it would be advantageous to put the complete 
           amount $\delta E$ in kinetic energy compared to potential energy.
     \item Effects of small deviations from the optimal angle are not large, since $x_M$ is stationary at the 
           optimal angle.
     \end{itemize}

\subsection{The effect of Air resistance}

Next we investigate the effect of the air resistance. This was considered in 
Tutevich (1969) \cite{Tutevich69}, Lichtenberg and Wills (1976) \cite{Lichtenberg78}. 
The effect of the air resistance is described by the following force 
\begin{equation}
F_W=m a_W\approx \frac{\varrho\pi c_w r^2v^2}{2},
\end{equation}
with the density of air $\varrho$, the drag coefficient of the sphere $c_w$ (about $1/2$),
the radius of the sphere $r$ and the velocity of the shot $v$. The maximum of $v_0$
in our calculations is about $16$ m/s which results in very small accelerations $a_W$. In addition
to the air resistance we included the wind velocity in our calculations.
\\
We confirm the results of  Tutevich (1969) \cite{Tutevich69}. 
As expected the effect of the air resistance turns out to be very small. Tutevich stated that 
for headwind with a velocity of $5$ m/s the shot is about $9-14$ cm
reduced for $v_0=12-14$ m/s compared to the value of $x_M$ without wind. He also stated that
for tailwind one will find an increased value of $6-8$ cm at $v_0=12-14$ m/s compared to
a windless environment. We could verify the calculations of Tutevich (1969) \cite{Tutevich69}
and obtain some additional information as listed in Table \ref{Tab2} by incorporating
these effects in a small Computer program that can be downloaded from the 
internet, see Rappl (2010)\cite{Rappl10}. An interesting fact is that headwind reduces the
shot more than direct wind from above (which could be seen as small $g$ factor corrections).
If the values of $v_0$, $h$ and $\theta$ are known (measured) precisely then the results of our program are exact.
\\
Now one can try to find again the optimal angle of release. Lichtenberg and Wills (1978) 
\cite{Lichtenberg78} find that the optimal angle is reduced compared to our previous 
determination by about $-0.13^\circ$ for some typical 
values of  $v_0$ and $h$ and by still assuming that  $v_0$, $h$ and $\theta$ are independent of each other.
\\
Due to the smallness of this effect compared to the remaining discrepancy of about $4^\circ$ between
the predicted optimal angle of release and the measurements we neglect air resistance in the following.
\section{Correlations between $\theta$, $v_0$ and $h$} 
Next we give up the assumption that the parameters $\theta$, $v_0$ and $h$
are independent variables. This was suggested e.g. in
Tricker and Tricker (1967) \cite{Tricker67}, 
Hay              (1973) \cite{Hay73}, 
Dyson            (1986) \cite{Dyson86},
Hubbard          (1988) \cite{Hubbard88},
de Mestre        (1990) \cite{deMestre90},
de Mestre et al. (1998) \cite{deMestre98},
Maheras          (1998) \cite{Maheras98},
Yeadon           (1998) \cite{Yeadon98},
Linthorne        (2001) \cite{Linthorne01}
and 
De Luca          (2005) \cite{deLuca05}.
We will include three effects: the dependence of the height of release from the angle of release,
the angle dependence of the force of the athlete and the effect of gravity during the delivery phase.
\subsection{The angle dependence of the point of release}
The height of the point, where the shot is released depends  
obviously on the arm length and on the angle
\begin{equation}
h = h_s + b \sin \theta,
\label{arm}
\end{equation}
with the height of the shoulder $h_s$ and the length of the arm $b$.
Clearly this effect will tend to enhance the value of the optimal
angle of release, since a larger angle will give a larger value 
of $h$ and this will result in a larger value of $x_M$.
This effect was studied in detail e.g. in de Luca (2005)\cite{deLuca05}.
We redid that analysis and confirm the main result of that work
\footnote{There was a misprint in Eq.(11) of \cite{deLuca05}: the term of order
$a^3$ should have a different sign.}.
The optimal angle can be expanded in $ a = h_s g/v_0^2$
\begin{eqnarray}
\sin \theta^{Opt.,deLuca} & = & 
\frac{1}{\sqrt{2}}
- \frac{  1 +   4 \sqrt{2}}{  16} a                               
+ \frac{160 + 113 \sqrt{2}}{ 512} a^2
- \frac{552 + 379 \sqrt{2}}{1024} a^3
\\
& \approx & 
\frac{1}{\sqrt{2}}
- 0.42  \frac{h_s g}{v_0^2}.
\end{eqnarray}
As expected above we can read off from this formula that the optimal angle of release
is now enhanced compared to the analysis of Lichtenberg and Wills (1976), 
\begin{equation}
\sin \theta^{Opt.,deLuca} - \sin \theta^{Opt.,Lichtenberg}
\approx   \frac{(0.35 h - 0.42 h_s) g}{v_0^2} > 0.
\end{equation}
For typical values of $v_0$, $h_s$ and $b$ de Luca (2005)\cite{deLuca05}
gets an increase of the optimal angle of release in the range of $+0.4^\circ$
to $+1^\circ$. With the inclusion of this effect the problem of 
predicting the optimal angle of release has become even more severe.
\subsection{The angle dependence of the force of the athlete}

The world records in bench-pressing are considerably higher than the
world records in clean and jerk. This hints to the fact that athletes
have typically most power at the angle $\theta = 0$ compared to larger values of $\theta$. 
This effect that is also confirmed by 
experience in weight training, was suggested and investigated e.g. by
McWatt    (1982)\cite{McWatt82},
McCoy     (1984)\cite{McCoy84},
Gregor    (1990)\cite{Gregor90}
and
Linthorne (2001)\cite{Linthorne01}.
The angle dependence of the force of the athlete can be measured and then be used
as an input in the theoretical investigation. Below we will use a very simple model
for the dependence to explain the consequences.
This effect now tends to favor smaller values for the optimal angle of release.

\subsection{The effect of gravity during the delivery phase}

Finally one has to take into account the fact, that the energy the athlete can produce
is split up in potential energy and in kinetic energy. 
\begin{eqnarray}
E & = & E_{kin} + E_{pot}
\\
  & = & \frac{1}{2} m v_0^2 + m g \delta h \, ,
\label{energy}
\end{eqnarray}
where $\delta h = h - h_s$ \footnote{We do not take into account an initial velocity 
of the shot, before the phase there the arm is stretched. In principle this can be 
simply included in our model - but we wanted to keep the number of parameters low.}.
Hence, the higher the athlete throws the lower will be the velocity of the shot. 
Since the achieved distance at shot put depends stronger on $v_0$ than on $h$ this
effect will also tend to giver smaller values for the optimal angle of release.
This was investigated e.g. in Tricker and Tricker (1967) \cite{Tricker67}, 
Zatsiorski and Matveev (1969) \cite{Zatsiorski69}, Hay (1973) \cite{Hay73} and 
Linthorne (2001)\cite{Linthorne01}.

\subsection{Putting things together}

Now we put all effects together. From Eq.(\ref{arm}) and Eq.(\ref{energy})
we get 
\begin{equation}
E =  \frac{1}{2} m v_0^2 + m g b \sin \theta.
\end{equation}
The angle dependence of the force of the athlete will result in an angle dependence
of the energy an athlete is able to transmit to the put
\begin{equation}
E = E (\theta) = E_0 f(\theta).
\end{equation}
The function $E(\theta)$ can in principle be determined by measurements with individual athletes.
From these two equations and from Eq.(\ref{arm}) we get
\begin{eqnarray}
v_0^2 (\theta) & = & 2 \left[ \frac{E(\theta)}{m} - g b \sin \theta \right],
\label{v0th}
\\
h     (\theta) & = & h_s + b \sin \theta       . 
\end{eqnarray}
Inserting these two $\theta$-dependent functions in Eq.(\ref{xM})
we get the full $\theta$-dependence of the maximum distance at shot put
\begin{equation}
x_M (\theta) = \frac{v_0^2(\theta) \cos \theta }{g} 
\left( \sin  \theta + \sqrt{\sin^2 \theta + \frac{2gh(\theta)}{v_0^2(\theta)}}
\right) \, .
\label{xMtheta}
\end{equation} 
The optimal angle of release is obtained as the root of the derivative of 
$x_M(\theta)$ with respect to $\theta$. To obtain numerical values for the optimal angle
we need to know $E(\theta)$. In principle this function is different for different
athletes and it can be determined from measurements with the athlete.
To make some general statements we present two simple toy models for $E (\theta)$.

\subsection{Simple toy models for $E(\theta)$}

We use the following two simple toy models for $E(\theta)$
\begin{eqnarray}
E_1 (\theta) & = & E_{1,0} \cdot f_1(\theta) = E_{1,0} \frac{2 + \cos \theta}{3}.
\\
E_2 (\theta) & = & E_{2,0} \cdot f_2(\theta) = E_{2,0} \left( 1- \frac{2}{3} \frac{\theta}{\pi} \right)
\end{eqnarray}
This choice results in $E = E_0$ for $\theta = 0$ and $E = \frac23 E_0$ for $\theta = \pi / 2$,
which looks reasonable. At this stage we want to remind the reader again: this Ansatz is just
supposed to be a toy model, a decisive analysis of the optimal angle of release will have to be done
with the measured values for $E(\theta)$.
We extract the normalization $E_0$ from measurements
\begin{eqnarray}
\frac{E_{1,0}}{m} & = & \frac{E_1(\theta)}{m f_1(\theta)} 
              = \frac{3}{2 + \cos \theta} \left[ g b \sin \theta + \frac{v_0^2}{2} \right],
\\
\frac{E_{2,0}}{m} & = & \frac{E_2(\theta)}{m f_2(\theta)} 
              = \frac{1}{1- \frac{2}{3} \frac{\theta}{\pi}} \left[ g b \sin \theta + \frac{v_0^2}{2} \right].
\end{eqnarray}
With the average values of Table  \ref{Tab1} ($h = 2.15$ m, $v_0 = 13.7$ m/s and $\theta = 38^\circ$)
and $b = 0.8$ m we get 
\begin{eqnarray}
h_s & = & 1.66 \, \mbox{m},
\\
\frac{E_{1,0}}{m} & = & 106.18 \, \mbox{m}^2/\mbox{s}^2,
\\
\frac{E_{2,0}}{m} & = & 103.305 \, \mbox{m}^2/\mbox{s}^2.
\end{eqnarray}
Now all parameters in $x_M(\theta)$ are known. Looking for the maximum of $x_M (\theta)$ 
we obtain the optimal angle of release to be 
\begin{eqnarray}
\theta_1^{Opt.} & = &  37.94^\circ \, , 
\\
\theta_2^{Opt.} & = &  37.38^\circ \, , 
\end{eqnarray}
which lie now perfectly in the measured range!
\\
Next we can also test the findings of Maheras (1998)\cite{Maheras98} that $v_0$ decreases linearly with
$\theta$ by plotting $v_0 (\theta)$ from Eq.(\ref{v0th}) against $\theta$. We find our toy
model 1 gives an almost linear decrease, while the decrease of toy model looks exactly linear.
\\
Our simple but reasonable toy models for the angle dependence of the force of the athlete
give us values for optimal release angle of about $37^\circ-38^\circ$, which coincide perfectly 
with the measured values. Moreover they predict the linear decrease of $v_0$ with increasing $\theta$ as
found by Maheras (1998)\cite{Maheras98} 
\section{Conclusion}
In this paper we have reinvestigated the biomechanics of shot put in order to determine the optimal angle of release.
Standard text-book wisdom tells us that the optimal angle is $\theta^{Opt.} = 45^\circ$, while measurements of 
top athletes tend to give values around $37^\circ - 38^\circ$.
Including the effect of the height of the athlete reduces the theory prediction for the optimal angle to 
values of about $42^\circ$ (Lichtenberg and Wills (1978) 
\cite{Lichtenberg78}).
As the next step we take the correlation between the initial velocity $\vec{v}_0$, 
the height of release $h$ and the angle of release $\theta$ into account.
Therefore we include three effects:
\begin{enumerate}
\item The dependence of the height of release from the angle of release is a simple geometrical relation.
       It was investigated in detail by de Luca (2005) \cite{deLuca05}. We confirm the result 
       and correct a misprint in the final formula of Luca (2005) \cite{deLuca05}.
       This effect favors larger values for the optimal angle of release.
\item The energy the athlete can transmit to the shot is split up in a kinetic part and a potential energy part.
       This effect favors smaller values for the optimal angle of release.
\item The force the athlete can exert to the shot depends also on the angle of release.
       This effect favors smaller values for the optimal angle of release.
\end{enumerate}
The third effect depends on the individual athlete. To make decisive statements the angle dependence
of the angular dependence of the force has to be measured first and then the formalism presented in 
this paper can be used to determine the optimal angle of release for an individual athlete.
\\
To make nethertheless some general statements we investigate two simple, reasonable toy models for the 
angle dependence of the force of an athlete.
With these toy models we obtain theoretical predictions for the optimal angle of $37^\circ - 38^\circ$,
which coincide exactly with the measured values. 
For our predictions we do not need initial measurements of $v(\theta)$ and $h(\theta)$ over a
wide range of release angles. In that respect our work represents a further developement of 
Linthorne (2001)\cite{Linthorne01}. Moreover our simple toy models predict the linear decrease 
of $v_0$ with increasing $\theta$ as found by Maheras (1998)\cite{Maheras98}.

\begin{acknowledgments}
We thank Philipp Weishaupt for enlightening discussions, Klaus Wirth for
providing us literature and J{\"u}rgen Rohrwild for pointing out a typo in one of our
formulas.
\end{acknowledgments}

\newpage

\section*{Tables}

\begin{table}[h]
	\centering
		\begin{tabular}{l|ccc|c||c|c}
		\hline Name & $v_0$ [m/s] & $h$ [m] & $\theta$ [$^\circ$] & $x_{\text{exp}}$ [m] & $x_M$ [m] & $\Delta x$ [m]\\\hline
		Woods & 13,9 & 2,2 & 40 & 21,17 & 21,61 & -0,44\\
		Woods & 13,7 & 2,1 & 35,7 & 21,05 & 20,58 & +0,47\\
		Woods & 13,6 & 2,16 & 37,7 & 20,88 & 20,59 & +0,29\\
		Briesenick & 14 & 2,2 & 39,7 & 21,02 & 21,87 & -0,85\\
		Feuerbach & 13,5 & 2,1 & 38,3 & 21,01 & 20,32 & +0,69
		\end{tabular}
	\caption{\label{Tab1}Compendium of some data measured during the Summer Olympic Games 1972 from Kuhlow\cite{Kuhlow75}}
\end{table}

\begin{table}[h]
	\centering
		\begin{tabular}{l||c|c|c|c|c}
		\hline Type & 1 m/s & 2 m/s & 3 m/s & 4 m/s & 5 m/s\\\hline
		Headwind $\Delta x$ [cm] & -1 & - 3 & -6 & -8 & -11\\
		Tailwind $\Delta x$ [cm] & 2 & 4 & 5 & 6 & 7\\
		From above $\Delta x$ [cm] & 0 & -1 & -2 & -3 & -4
		\end{tabular}
	\caption{\label{Tab2}Differences for $x_M$ with $v_0=13$ m/s, $\theta=45^\circ$ and $h=2$ m compared to zero wind velocity calculated by Rappl (2010) \cite{Rappl10}}
\end{table}

\newpage

\section*{Figures}

\begin{figure}[h]
\begin{center}
\scalebox{1.40}{\includegraphics{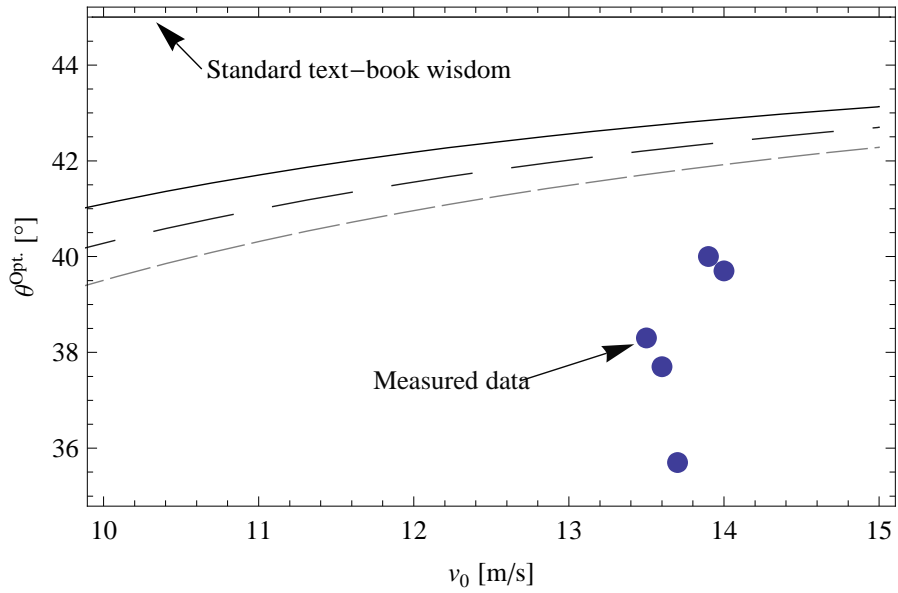}}
\caption{\label{Fig3}The optimal angle $\theta^{Opt.}$ in the interesting area of $v_0$ between $10-15$ m/s for (from top to bottom) $h=1.6,2.0,2.4$ m taking the height of the athlete into account and the measured data from Table \ref{Tab1} shown as dots}
\end{center}
\end{figure}

\end{document}